\pdfoutput=1

\documentclass[10pt,onecolumn,twoside]{IEEEtran}

\IEEEoverridecommandlockouts                              
\usepackage{enumerate}
\usepackage{url}
\usepackage{graphicx}
\usepackage[table]{xcolor}
\usepackage{booktabs}
\usepackage{verbatim}
\makeatletter
\newcommand{\verbatimfont}[1]{\def\verbatim@font{#1}}%
\makeatother
\verbatimfont{\ttfamily\small}
\usepackage{multirow}
\usepackage{array}
\usepackage{amsmath}

\usepackage{amsthm}
\usepackage{tabu}
\usepackage{amssymb}
\usepackage{graphics}
\usepackage{mathrsfs}
\usepackage{cite}
\usepackage[mathscr]{euscript}
\usepackage{nameref}

\newcommand{\SSL}{{\rm SS}}
\newcommand{\PNE}{{\rm PNE}}
\newcommand{\ABR}{{\rm ABR}}
\newcommand{\abr}{{\rm br}}
\newcommand{\aabr}{a^{\abr}}
\newcommand{\pabr}{P^{\abr}}

\newcommand{\vs}{\vspace{-1mm}}
\newcommand{\hs}[1][0.5]{\hspace{-#1mm}}
\newcommand{\pone}{Player~$1$} 
\newcommand{\ptwo}{Player~$2$}
\newcommand{\pthr}{Player~$3$}

\newcommand{\aaa}{\mathcal{A}}
\newcommand{\ep}{\epsilon}
\DeclareMathOperator*{\argmax}{arg\,max}

\newcommand{\gee}{\mathcal{G}}

\newtheorem{theorem}{Theorem}
\newtheorem{prop}[theorem]{Proposition}

\newtheorem{corollary}[theorem]{Corollary}
\newtheorem{definition}[theorem]{Definition}

\newcommand{\shd}{\cellcolor{gray!50}}

\title{Are Multiagent Systems Resilient to Communication Failures?
}
\author{{Philip N. Brown, Holly P. Borowski, and Jason R. Marden}
\thanks{A preliminary abridged version of this work was submitted to the 2018 American Control Conference.}
\thanks{This work is supported ONR Grant \#N00014-17-1-2060 and NSF Grant \#ECCS-1638214.}
\thanks{P. N. Brown is a PhD candidate with the Department of Electrical and Computer Engineering, University of California, Santa Barbara, CA, {\texttt{pnbrown@ece.ucsb.edu}}, corresponding author.}  
\thanks{H. P. Borowski is a research scientist with Numerica Corporation, Fort Collins, CO, {\texttt{hollyboro@gmail.com}}.}
\thanks{J. R. Marden is an associate professor with the Department of Electrical and Computer Engineering, University of California, Santa Barbara, CA, {\texttt{jrmarden@ece.ucsb.edu}}.}
}

\begin{document}

\maketitle

\begin{abstract}
A challenge in multiagent control systems is to ensure that they are appropriately resilient to communication failures between the various agents.
In many common game-theoretic formulations of these types of systems, it is implicitly assumed that all agents have access to as much information about other agents' actions as needed.
This paper endeavors to augment these game-theoretic methods with policies that would allow agents to react on-the-fly to losses of this information.
Unfortunately, we show that even if a single agent loses communication with one other weakly-coupled agent, this can cause arbitrarily-bad system states to emerge as various solution concepts of an associated game, regardless of how the agent accounts for the communication failure and regardless of how weakly coupled the agents are.
Nonetheless, we show that the harm that communication failures can cause is limited by the structure of the problem; when agents' action spaces are richer, problems are more susceptible to these types of pathologies.
Finally, we undertake an initial study into how a system designer might prevent these pathologies, and explore a few limited settings in which communication failures cannot cause harm.

\end{abstract}




%
\vs
\section{Introduction}


Many control systems are intrinsically distributed, with actuators, sensors, and decision-makers not physically co-located.
Network routing, power generation, and autonomous vehicles are immediate examples~\cite{Jadbabaie2003,Brown2017a,Banerjee2015}.
If centralized control architectures are infeasible due to communication or computation constraints, this may prompt designers to employ a distributed decision-making architecture in which control decisions are made locally by each agent as a function of locally-available information~\cite{Martinez2007,Nedic2009}.

In a distributed multiagent system, it can be helpful to think of the design and operation phases as distinct from one another; we call these phases ``offline'' and ``online,'' respectively.
In the offline phase, the system designer endows each agent with a set of local decision-making rules that will be used to solve an as-yet unspecified problem.
Then, in the online phase, the details of the problem are realized and the agents each execute their predetermined rules with no more input from the designer.
If an unplanned communication failure occurs which reduces the amount of information available to some agent, that agent must react to the situation on-the-fly, as the system designer is no longer available to make further changes to the control algorithm.
The key question here is this: in the offline phase, how can agents be endowed with policies to react to online losses of important pieces of information, such that these policies lead to desirable emergent collective behavior?

In a typical game-theoretic approach, during the offline phase, each agent is assigned a utility function (i.e., local objective function) derived from the system objective, and endowed with a learning algorithm by which it will attempt to maximize its own utility function during the online phase~\cite{Li2013,Borowski2013,Marden2014,Gopalakrishnan2014}.
A variety of results show that if utility functions are designed properly, there exist local learning algorithms that are guaranteed to converge to efficient Nash equilibria in several game classes of interest (e.g., potential games), even when the details of the game instance are not known in the offline phase~\cite{Lim2013,Marden2013a,Candogan2013,Ali2016a}.

These learning algorithms are typically designed to be uncoupled, meaning that agents do not know the utility functions of other agents~\cite{Hart2003,Fudenberg1998}.
Some of these algorithms (e.g., the logit response dynamics~\cite{Alos-Ferrer2010}) essentially require each agent to know the current action choices of all other agents, as well as the exact form of their own utility functions.
When this is the case, it is assumed that information about other agents' actions is freely available, ostensibly either via communication or direct observation~\cite{Marden2012}.
Alternatively, in so-called payoff-based algorithms, at each time step an agent merely realizes the output of its own utility function as a ``payoff,'' and attempts to play actions that are historically correlated with high payoffs~\cite{Kandori1993,Marden2009,Marden2009a}.
Here, explicit communication between the agents is avoided by the assumption that agents merely realize payoffs, but the payoffs themselves are essentially provided by an unspecified oracle.
Thus, these approaches do not adequately capture a situation in which an agent must react online to the complete loss of information about the action choices of some other relevant agent.

If communication between agents is unreliable, it would be desirable to endow agents with policies which allow them to update their utility functions online in response to losing information about other agents' actions.
For concreteness, consider the following $3$-player, $2$-action game, for some small $\delta>0$.
Note that we use ``players'' and ``agents'' interchangeably.

\setlength{\extrarowheight}{2pt}
\begin{center}
\begin{tabular}{cc|c|c|cc|c|c|}
      & \multicolumn{1}{c}{} & \multicolumn{2}{c}{Player $2$} 					& \multicolumn{2}{c}{} & \multicolumn{2}{c}{Player $2$}\\
      & \multicolumn{1}{c}{} & \multicolumn{1}{c}{$A$}  & \multicolumn{1}{c}{$B$} 	& \multicolumn{2}{c}{} & \multicolumn{1}{c}{$A$} & \multicolumn{1}{c}{$B$}  \\\cline{3-4}\cline{7-8}
      \multirow{2}*{Player $1$}  
      & $A$ 		& $\shd1-\delta,\ 1$ 	& $0,\ 0$ 		& \multicolumn{2}{r|}{$\hspace{.5in} A$}		& $0,\ 2\delta$  	& $2\delta,\ 0$	\\\cline{3-4}\cline{7-8}
      & $B$ 		& $\delta,\ 0$ 	& $1,\ 0$ 		& \multicolumn{2}{r|}{$B$}		&$3\delta,\ \delta$ 	& $\delta,\ 2\delta$		\\\cline{3-4}\cline{7-8}
      \multicolumn{2}{l}{Player $3$} & \multicolumn{2}{c}{$A$} & \multicolumn{2}{c}{} &\multicolumn{2}{c}{$B$}
\end{tabular}
\end{center}

The rows indicate the action choices available to \pone, the columns indicate the action choices available to \ptwo, and \pthr\ selects between the left and right matrices.
Let \ptwo\ and \pthr\ have identical utility functions, so each cell in the matrices depicts the payoffs for \pone\ and \ptwo/3, respectively as a function of joint action choices.

The optimal action profile (measured by the sum of the players' payoffs) is the upper-left, and all best-response paths lead to this unique Nash equilibrium.%
\footnote{
A best-response path is a sequence of unilateral payoff-maximizing action updates by players, formally defined in Section~\ref{sec:cert}; a Nash equilibrium is a joint action profile from which no agent can unilaterally deviate and improve payoff, defined in~\eqref{eq:pnedef}.
}
A salient question here is this: if \pone\ does not know the action of \ptwo\ (and also does not know \ptwo's utility function), how should \pone\ evaluate her own action choices?
In this paper we assume that \pone\ knows what payoffs she \emph{could} receive for any action choice by \ptwo; for example, If \pone\ and \pthr\ are both playing $A$, \pone\ knows that her payoff is either $1-\delta$ or $0$.
Given this information, \pone\ needs to assign a ``proxy payoff'' to this situation as a function of the payoffs she could be receiving; we call such a policy for computing proxy payoffs an \emph{evaluator}, which we formally define in Definition~\ref{def:eval} (Section~\ref{sec:model}).
One simple evaluator is to choose the maximum payoff from each row, yielding this effective payoff matrix:
%
\setlength{\extrarowheight}{2pt}
\begin{center}
\begin{tabular}{cc|c|c|cc|c|c|}
      & \multicolumn{1}{c}{} & \multicolumn{2}{c}{Player $2$} 					& \multicolumn{2}{c}{} & \multicolumn{2}{c}{Player $2$}\\
      & \multicolumn{1}{c}{} & \multicolumn{1}{c}{$A$}  & \multicolumn{1}{c}{$B$} 	& \multicolumn{2}{c}{} & \multicolumn{1}{c}{$A$} & \multicolumn{1}{c}{$B$}  \\\cline{3-4}\cline{7-8}
      \multirow{2}*{Player $1$}  
      & $A$ 		& $\hs1-\delta,\ 1\hs$ 	& $\hs1-\delta,\ 0\hs$ 		& \multicolumn{2}{r|}{$\hspace{0.5in} A$}		& $2\delta,\ 2\delta$  	& $2\delta,\ 0$	\\\cline{3-4}\cline{7-8}
      & $B$ 		& $1,\ 0$ 	& $1,\ 0$ 		& \multicolumn{2}{r|}{$B$}		&$3\delta,\ \delta$ 	& $\shd\hs 3\delta,\ 2\delta\hs$		\\\cline{3-4}\cline{7-8}
      \multicolumn{2}{l}{Player $3$} & \multicolumn{2}{c}{$A$} & \multicolumn{2}{c}{} &\multicolumn{2}{c}{$B$}
\end{tabular}
\end{center}

%
%
%

Note that \ptwo\ and \pthr's payoffs are unchanged, but \pone's perceived payoffs are now independent of \ptwo's action.
In this modified game, \pone\  now always prefers action $B$ over action $A$, and this causes all best-response paths to lead to the inefficient lower-right action profile.
Furthermore, note that this conclusion stands for any evaluator chosen by \pone: if she had chosen the minimum payoff in each row rather than the maximum, the result is unchanged -- and the same holds for mean and sum.
This suggests that communication failures can cause catastrophic problems in multiagent systems.

However, it appears clear from the nominal payoffs that \ptwo\ was important to \pone's decision-making, and this made it impossible for \pone\ to compute helpful proxy payoffs when information about \ptwo\ was lost.
However, if \ptwo\ had been ``inconsequential'' to \pone\ in some sense, 
would that have lessened the harm of the communication failure?
Our goal is to explore this issue and determine when and how it is possible to endow agents with policies for computing proxy payoffs to protect against losses of information about weakly-coupled agents.
Specifically, some questions addressed in this paper are
\begin{enumerate}
\item What is a meaningful notion of ``weak coupling?''
\item Is there a method of computing proxy payoffs which can provide some guarantee that emergent behavior is good despite lost information?
\item Are there particular problem structures that confer greater degrees of resilience than others?
\end{enumerate}

This paper's main contribution is to show that \emph{regardless of how proxy payoffs are chosen,} in several  settings, loss of information about even weakly-coupled agents can cause arbitrarily-low-quality states to emerge as various game solution concepts.
Definition~\ref{def:inconsequential} formalizes a notion of weak coupling that we term \emph{inconsequentiality}: if \ptwo\ is $\ep$-inconsequential to \pone\ this means that for any joint action, a unilateral action change by \ptwo\ can cause no more than an $\ep$ change in \pone's payoff (See Definition~\ref{def:inconsequential} in Section~\ref{sec:pg}).
That is, even if \ptwo's action is unknown, \pone\ can always estimate her own payoff to within $\ep$.

In Section~\ref{sec:pg}, we consider the well-studied class of potential games~\cite{Monderer1996}.
Theorem~\ref{thm:pgbad} shows that there exist potential games such that \emph{for every reasonable payoff evaluator,} if one agent loses information about a single inconsequential agent's action, all of a game's Nash equilibria can be made arbitrarily bad.
Next, we study the effect of noisy learning dynamics on these types of problems.
Stochastic learning rules such as log-linear learning are known to stabilize high-quality states of games in many situations~\cite{Alos-Ferrer2010,Marden2012}.
However, our Theorem~\ref{thm:sspg} shows that the pathologies shown in Theorem~\ref{thm:pgbad} can also occur when agents update their actions according to log-linear learning.

Subsequently, in Section~\ref{sec:ii} we show in Proposition~\ref{prop:iinotasbad} that the class of identical interest games is slightly better-behaved than general potential games, but Theorem~\ref{thm:ss} shows that noisy learning dynamics can actually destabilize efficient Nash equilibria and again, there is no general way to compute proxy payoffs that can prevent inefficient action profiles from emerging as stochastically stable states of log-linear learning.

Section~\ref{sec:positive} presents a pair of positive results showing features which can limit the harm of communication failures when agents update their actions stochastically according to log-linear learning.
Theorem~\ref{thm:all} shows that if \emph{every} agent loses information about a single inconsequential agent's action, identical-interest games retain their desirable properties and high-quality states remain stochastically stable.
Furthermore, we show in Theorem~\ref{thm:candogan} that the damage a communication failure can cause in a potential game is limited by the total number of action profiles in a game.

Lastly, Section~\ref{sec:cert} proposes a way to certify whether a set of proxy payoffs is susceptible to the types of pathologies demonstrated in this paper.
This certificate looks for some notion of alignment between the chosen proxy payoffs and the nominal potential function of the game; if these are sufficiently aligned, then losing communication with inconsequential players can cause no harm.

\section{Model} \label{sec:model}

\subsection{Game theoretic preliminaries}

We model the multiagent system as a finite strategic-form game with player set $\mathcal{I} =\left\{1,\ldots,n\right\}$ where each player $i\in\mathcal{I}$ selects an action from action set $\aaa_i$.
The preferences of each player over his actions are encoded by a utility function $U_i:\aaa\rightarrow\mathbb{R}$ where $\aaa = \aaa_1\times\cdots\times\aaa_n$; that is, each player's payoff is a function of his own action and the actions of all other players.
A game is specified by the tuple $G=\left(\mathcal{I},\aaa,\{U_i\}_{i\in\mathcal{I}}\right)$.

For an action profile $a=(a_1,a_2,\ldots,a_n)\in\aaa$, let $a_{-i}$ denote the profile of player actions other than player $i$; i.e.,
\vs
\begin{equation*}
a_{-i} = (a_1,\ldots,a_{i-1},a_{i+1},\ldots,a_n).
\vs
\end{equation*}
Similarly, $a_{-ij}$ denotes the profile of player actions other than players $i$ and $j$:
\vs
\begin{equation*}
a_{-ij} = (a_1,\ldots,a_{i-1},a_{i+1},\ldots,a_{j-1},a_{j+1},\ldots,a_n).
\vs
\end{equation*}
With this notation, we will sometimes write an action profile $a$ as $(a_i,a_{-i})$.
Similarly, we may write $U_i(a)$ as $U_i(a_i,a_{-i})$.
Let $\aaa_{-i}=\Pi_{j\neq i}\aaa_j$ denote the set of possible collective actions of all players other than player $i$.
We define player $i$'s \emph{best response set} for an action profile $a_{-i}\in\aaa_{-i}$ as
$B_i(a_{-i}) :=  \argmax_{a_i\in\aaa_i}\ U_i\left(a_i,a_{-i}\right).$
%
An action profile $a^{\rm ne}\in\aaa$ is known as a \emph{pure Nash equilibrium} if for each player $i$,
\vs
\begin{equation}
a^{\rm ne}_i \in B_i\left(a^{\rm ne}_{-i}\right). \label{eq:pnedef}
\vs
\end{equation}
That is, all players are best-responding to each other.
The set of pure Nash equilibria of game $G$ is denoted ${\rm PNE}(G)$.

\subsection{Distributed optimization and game classes}

We assume that there is a global welfare function (i.e., objective function) $W:\aaa\rightarrow[0,1]$ that the players are trying to maximize.
We assume without loss of generality that at least one action profile achieves the welfare maximum; i.e., in every game there exists $\bar{a}\in\aaa$ such that $W(\bar{a})=1$.%

A common game-theoretic formulation of distributed optimization problems involves assigning utility functions to players that are derived from $W$ in some way that ensures that $\argmax_{a}W(a)\subseteq{\rm PNE}(G)$. 
One way to assign utility functions that accomplishes this is simply to give each player~$i$ the full objective function, or let 
$$U_i(a_i,a_{-i}):=W(a_i,a_{-i}).$$
This induces an \emph{identical-interest game}, or one in which all players have the same utility function.
In this paper, if a game is stated to be an identical-interest game, it is assumed that the common utility function is the welfare function $W$.

A somewhat more nuanced method of assigning player utility functions is known as ``marginal-contribution'' utility design~\cite{Wolpert1999a}.
Here, for each player~$i$, arbitrarily select a baseline action $a_i^b\in\aaa_i$, and assign the player a utility function of
$$
U_i(a_i,a_{-i}) := W(a_i,a_{-i}) - W(a^b_i,a_{-i}).
$$
Assigning utility functions in this way ensures that in the resulting game, whenever a player unilaterally changes actions in a way that improves her utility, this improves $W$ by the same amount.
Formally, for every player $i\in\mathcal{I}$, for every $a_{-i}\in\aaa_{-i}$, and for every $a'_i,a''_i\in\aaa_i$,
\begin{equation}
U_i\left(a_i',a_{-i}\right) - U_i\left(a_i'',a_{-i}\right) = W\left(a_i',a_{-i}\right) - W\left(a_i'',a_{-i}\right). \label{eq:pg}
\end{equation}

A game that satisfies~\eqref{eq:pg} for some $W$ is known as a \emph{potential game} with \emph{potential function} $W$~\cite{Monderer1996}.
In a potential game, there is a strong notion of alignment between the interests of the various players, and action profiles which maximize $W$ are always pure Nash equilibria: $\argmax_{a}W(a)\subseteq{\rm PNE}(G)$.
Further, many distributed learning algorithms converge in potential games to potential-maximizing action profiles~\cite{Marden2013}.

Throughout this paper, if a game is stated to be a potential game, it is assumed that its potential function is equal to the welfare function $W$.
For non-potential games, the welfare function will be specified as needed.

\subsection{Online learning and its associated solution concepts}

Once the utility functions have been assigned offline, how should the agents choose their actions online?
In this paper, we have agents update their actions by stochastic asynchronous  processes that proceed as follows.
Starting at some initial joint action $a(0)$, at each time $t\in\mathbb{N}$ an agent $i$ (called the \emph{updating agent}) is selected uniformly at random from $\mathcal{I}$ to choose an action $a_i(t+1)$ to play at time $t+1$, and all other agents simply repeat their previous action: $a_{-i}(t+1)=a_{-i}(t)$.

\subsubsection*{Asynchronous best reply process} 
The prototypical learning rule that we will subsequently build upon is the \emph{asynchronous best reply process}, where at each time step the updating agent $i$ selects an action from their best response set uniformly at random.
Formally, conditional on agent $i$ being the updating agent at time $t$, the probability that agent $i$ selects action $a'_i$ in the next time step is given by
\begin{equation}
{\rm Pr}\left[a_i(t+1)=a'_i \ |\  a_{-i}(t)\right] = \left\{
\begin{array}{cl}
\frac{1}{\left|B_i(a_{-i}(t)) \right|} 	& \mbox{ if }a'_i \in B_i(a_{-i}(t)), \\
0 							& \mbox{ otherwise.}
\end{array}
\right.  \label{eq:abr}
\end{equation}

For any strategic-form game, the asynchronous best reply process defines a Markov process $\pabr$ over joint action profiles $\aaa$; as such, we will frequently refer to a joint action profile $a$ as a ``state'' of $\pabr,$ and denote a sample path of $\pabr$ as $\aabr(t)$.
A \emph{recurrent class} of $\pabr$ is a set of action profiles $A\subseteq\aaa$ such if the process is started in $A$ it remains in $A$ (i.e., if $\aabr(0)\in A$, then $\aabr(t)\in A$ for all $t>0$), and for any states $a,a'\in A$, the probability that the process started at $a$ eventually visits $a'$ is positive: ${\rm Pr}\left[\left. \aabr(t)=a' \mbox{ for some }t\geq1 \right| \aabr(0)=a \right] > 0.$
We denote the set of recurrent classes of the asynchronous best reply process of a game $G$ under $\pabr$ by $\ABR(G)$; since the state space is finite, $\ABR(G)$ is never empty.
With a slight abuse of notation, we will occasionally write $a\in\ABR(G)$ to mean that $a\in A$ for some $A\in\ABR(G)$.

The recurrent classes $\ABR(G)$ fully characterize the long-run behavior of $\pabr$ in the sense that for any initial joint action $\aabr(0)$, standard Markov results show that the process eventually enters some recurrent class almost surely.
In general, a game may have many recurrent classes, but if $G$ is a potential game with a unique pure Nash equilibrium $a^{\rm ne}$, then it holds that $\{a^{\rm ne}\}$ is the unique recurrent class of $\pabr$ and thus $\aabr(t)$ converges to $a^{\rm ne}$ almost surely.
However, even a potential game may have multiple strict Nash equilibria, in which case each equilibrium forms its own recurrent class -- and it is not possible to predict \emph{a priori} which of these will eventually capture the process.
Moreover, it is possible for $\pabr$ to have a cycle as a recurrent class.

\subsubsection*{Log-linear learning preliminaries}

Motivated in part by this ambiguity, a rich body of literature has developed to study ``noisy'' best-reply processes, in which a nominal best-reply process is perturbed by a small amount of noise to make its associated Markov chain ergodic.
The typical setup is that players are randomly offered opportunities to choose a new action, and with some small probability, they choose suboptimal actions.
In the context of a social system, this ``choice of suboptimal action'' is viewed as the player mistakenly choosing an action that is not a best response; if the game is meant to model an engineered system, the suboptimal action is used as a means for the agents to explore the state space~\cite{Pradelski2012,Blume1993,Blume1995,Young1993,Marden2012,Tatarenko2014}.
One such noisy update process is the \emph{log-linear learning} rule, defined as follows.%
\footnote{
This paper investigates log-linear learning in particular for reasons of parsimony and concreteness; several of our results can be readily extended  to more general noisy best-response processes.
Formally making this extension would considerably increase the notational density of the paper while adding little substance.
}

At time $t$, updating agent $i$ chooses its next action probabilistically as a function of the payoffs associated with the current action profile: the probability of choosing some action $a'_i$ in the next time step is given by
\begin{equation}
{\rm Pr}\left[a_i(t+1)=a'_i \ |\  a_{-i}(t)\right] = \frac{e^{\beta U_i(a'_i,a_{-i}(t))}}{\sum_{a_i\in\aaa_i}e^{\beta U_i(a_i,a_{-i}(t))}}, \nonumber
\end{equation}
where $\beta>0$ is a parameter indicating the degree to which agents desire to select their best response.
If $\beta=0$, agents select actions uniformly at random; as $\beta\rightarrow\infty$, the limiting process is the asynchronous best-reply process of Section~\ref{sec:model}.

For any $\beta>0$, log-linear learning induces an ergodic Markov process on $\aaa$; denote its unique stationary distribution by $\pi^\beta$.
It is well-known that the limiting distribution $\pi\triangleq\lim_{\beta\rightarrow\infty}\pi^\beta$ exists, and that it is a stationary distribution of the asynchronous best-reply process.
Let $\pi(a)$ denote the probability of joint action $a\in\aaa$ being played in the limiting distribution $\pi$; if $\pi(a)>0$, then action profile $a$ is said to be \emph{stochastically stable} under log-linear learning.
It is known that for any game $G$,
\begin{equation}
a\in\SSL(G) \ \implies \ a\in\ABR(G). \label{eq:ssinabr}
\end{equation}
Furthermore, existing results give us that in a potential game with potential function $W$, the set of stochastically stable action profiles under log-linear learning is equal to the set of potential function maximizers:
\begin{equation} \label{eq:potmaxss}
\SSL(G) = \argmax\limits_{a\in\aaa}W(a).
\end{equation}

\subsection{Modeling communication failures} \label{ssec:commfail}

To model the effect of a communication failure, let a single player lose access to information about the action of a single other player; without loss of generality, let these players be \pone\ and \ptwo, respectively.
We will commonly say that ``\ptwo\ is \emph{hidden} from \pone.''
Note that more general formulations are possible, but in this paper we investigate this special case as it suffices to highlight several important issues. Furthermore, it seems likely that allowing more than one communication failure in a game would worsen our results.

When \pone\ cannot observe \ptwo's action, this means that the utility function $U_1$ must be modified in some way so that it no longer depends on the action choice of \ptwo.
That is, \pone\ must adopt a \emph{proxy payoff function} $\tilde{U}_1(a_1,a_{-12})$ that becomes the new basis for decision-making, but that does not depend on the action choice of \ptwo.

Computing $\tilde{U}_1$ means assigning a value to each $(a_1,a_{-12})$, taking into account that the true utility is a function of the unobservable action of \ptwo\ as well. That is, the true utility is some unknown number in the set
$\left\{U_1(a_1,a_2,a_{-12}) :  a_{2}\in\aaa_2  \right\}.$
To compute proxy payoffs, we assign \pone\ an \emph{evaluator} $f$, which is a function that for each $(a_1,a_{-12})$, takes the set of possible payoffs and returns a proxy payoff:
\begin{equation}\label{eq:evaldef}
\tilde{U}_1\left(a_1,a_{-12})\right) = f\left(\left\{U_1(a_1,a_2,a_{-12}) :  a_{2}\in\aaa_2  \right\}\right).
\end{equation}
The space of feasible evaluators is large, and we restrict it in only two simple ways, as indicated by the following definition:
\begin{definition} \label{def:eval}
An \emph{acceptable evaluator} $f$ is a mapping from sets of numbers to $\mathbb{R}$ satisfying the following properties.
Let $S=(s_i)_{i=1}^{k}$ and $S'=(s'_i)_{i=1}^{k}$, where $S$ and $S'$ are assumed to be ordered increasing:
\begin{enumerate}
\item If $s_i>s'_i$ for each $i$, then $f(S)>f(S')$,
\item If $s_i=s'_i$ for each $i$, then $f(S)=f(S')$.
\end{enumerate}
If an acceptable evaluator further satisfies $f(S)\in \left[\min(S),\max(S)\right]$, it is called a \emph{bounded acceptable evaluator}.
\end{definition}

If \pone\ uses acceptable evaluator $f$ to compute proxy payoffs for the case when \ptwo's action is unobservable, we say that \pone\ \emph{applies} $f$ to \ptwo.
Proposition~\ref{prop:eval} gives a partial list of evaluators which are acceptable by Definition~\ref{def:eval}; its proof is included in the Appendix.

\begin{prop} \label{prop:eval}
The following functions are acceptable evaluators; numbers (2) through (4) are also bounded:
\begin{enumerate}
\item Sum: $f_{\rm sum}(S) = \sum_{s\in S} s$
\item Maximum element: $f_{\rm max}(S) = \max_{s\in S} s$
\item Minimum element: $f_{\rm min}(S) = \min_{s\in S} s$
\item Mean: $f_{\rm mean}(S) = \frac{1}{|S|}\sum_{s\in S} s$
\end{enumerate}
\end{prop}

Given a nominal game $G$ and evaluator $f$, we write $G_{f}$ to denote the \emph{reduced} game generated when \pone\ applies evaluator $f$ to \ptwo.

\subsection{Assessing the quality of an evaluator}

We consider an acceptable evaluator $f$ to be effective if for any nominal game $G$, the reduced game $G_f$ induced by $f$ is similar to the nominal game, where this similarity is measured by the welfare of the games' equilibria.
For a game $G$, let $\mathcal{E}(G)$ be some set of equilibria associated with $G$; for example, $\mathcal{E}(G)$ could represent the set of recurrent classes of the asynchronous best-reply process for $G$.
In the forthcoming, we write $\gee$ to denote a given class of games.

This paper presents a number of negative and positive results; to show the negative results, we use an optimistic measure of quality given by
\begin{equation} \label{eq:Qopt}
Q_\mathcal{E}^-\left(\gee,f\right) \triangleq\inf_{G\in\gee} 
\frac{
	\max\limits_{a\in\mathcal{E}(G_{f})}W(a)}
	{\min\limits_{a\in\mathcal{E}(G)}W(a)}.
\end{equation}
Note that here, by checking the ratio of maximum to minimum welfare, this quality measure is designed to produce the highest values possible.
If~\eqref{eq:Qopt} is close to $0$, this indicates that for some game $G\in\gee$, the \emph{best} equilibria induced by $f$ can perform far worse than the \emph{worst} equilibria of the nominal game.
Accordingly, all of our negative results show situations in which this form of the quality metric can be close to $0$.
We sometimes wish to evaluate~\eqref{eq:Qopt} on an individual game (i.e., a singleton class of games) and denote this with $Q^-_\mathcal{E}(G,f):=Q^-_\mathcal{E}\left(\{G\},f\right)$.

Alternatively, to show the positive results, we use a pessimistic measure of quality given by
\begin{equation}\label{eq:Qpess}
Q_\mathcal{E}^+\left(\gee,f\right) \triangleq\inf_{G\in\gee} 
\frac{
	\min\limits_{a\in\mathcal{E}(G_{f})}W(a)}
	{\max\limits_{a\in\mathcal{E}(G)}W(a)}. 
\end{equation}
In contrast to~\eqref{eq:Qopt}, here by checking the ratio of minimum to maximum welfare, the quality measure is designed to produce the lowest values possible.
That is, if~\eqref{eq:Qpess} is close to $1$, this indicates that for every game in $\gee$, the \emph{worst} equilibria induced by $f$ are nearly as good as the \emph{best} equilibria of the nominal game.
Accordingly, all of our positive results show situations in which this form of the quality metric can be close to $1$.

We can now state the main goal of this paper.
We wish to find payoff evaluators $f$ which can ensure that these measures of quality are high for meaningful classes of games.
That is, given some $\gee$ and $\mathcal{E}$, we wish to find $f$ to maximize~$Q_\mathcal{E}^+\left(\gee,f\right)$.

\section{Resilience against communication failures is challenging}

In multiagent systems, it is an attractive goal to endow agents with local policies which allow them to react on-the-fly to losses of information about other agents' actions.
Unfortunately, the results in this section suggest that there exist no general uncoupled methods for doing so.
Unless otherwise stated, proofs of all theorems appear in the Appendix.


First, note that if no restriction is placed on which types of game we are considering, a single communication failure can easily and catastrophically degrade performance.
The following proposition assumes that \pone\ loses information about the action choice of \ptwo, and thus must compute proxy payoffs for the case when \ptwo's action is unobservable.

\begin{prop} \label{prop:bad}
Let $\gee$ be the set of all games.
If Player~$1$ applies any acceptable evaluator $f$ to Player~$2$, then for all $\ep>0$, it holds that
\begin{equation}
Q_\ABR^-\left(\gee,f\right)  \leq \ep. \label{eq:pgbad}
\end{equation}
\end{prop}
Here, by showing that the optimistic measure $Q_\ABR^-(\gee,f)$ is close to $0$, we see that in general games, no evaluator can prevent best-response processes from selecting arbitrarily-inefficient action profiles.

\begin{proof}
This is shown using the game presented in the introduction paired with welfare function $W(a) = \sum_i U_i(a)$ (normalized to have a range of $[0,1]$).
In that game, if \ptwo\ is hidden from \pone\ there exists no acceptable evaluator which can prevent the bottom-right action profile from being the unique strict Nash equilibrium to which all best-response paths lead.
This is because for any acceptable evaluator $f$, action $B$ is a strictly dominant strategy  for \pone.
The upper-left action profile has welfare $3-\delta$, but the lower-right action profile has welfare $5\delta$.
Thus, for any $\ep>0$, letting $0<\delta<3\ep/(5+\ep)$ yields the proof.
\end{proof}

\section{Are potential games resilient to communication failures?} \label{sec:pg}

\subsection{Results for recurrent classes of the asynchronous best reply process}

It is clear from Proposition~\ref{prop:bad} that some structure is needed in order to prevent communication failures from causing harm.
Potential games offer a natural starting point for studying resilience in this context; intuitively, since the payoffs of agents in a potential game are nicely aligned, this might offer a degree of protection.


Furthermore, we wish to study games in which players are in some sense only weakly coupled to the players they cannot observe.
We introduce the following notion of weak interrelation: we say Player~$2$ is ``inconsequential'' to Player~$1$ if \ptwo\ can never cause a large change in \pone's payoff by changing actions.
We make this notion precise in Definition~\ref{def:inconsequential}:

\begin{definition} \label{def:inconsequential}
Player $j$ is $\epsilon$-\emph{inconsequential} to player $i$ if for all $a_i\in\aaa_i$, all $a_{-ij}\in\aaa_{-ij}$, and all $a_j,a_j'\in\aaa_j$,
\begin{equation}
\left| U_i\left(a_i,a_j,a_{-ij}\right) - U_i\left(a_i,a_j',a_{-ij}\right)\right| \leq \ep.
\end{equation}
\end{definition}

Now, let $\gee^{\rm PG}_\ep$ denote the class of potential games for which \ptwo\ is no more than $\ep$-inconsequential to \pone.
Our standing assumption will be that in each game, \pone\ loses information about the action of \ptwo, and thus must apply an acceptable evaluator to \ptwo.
That is, for each game $G\in\gee^{\rm PG}_\ep$, we are assured that even if \pone\ cannot observe \ptwo's action, a unilateral deviation by \ptwo\  can have only a small impact on \pone's payoff.
One might hope that by imposing the additional structure provided by potential games and inconsequentiality that the severe pathologies of Proposition~\ref{prop:bad} could be avoided.
Indeed, it can be readily shown that for the special case of $\ep=0$, for any game $G\in\gee_{\ep=0}^{\rm PG}$, if \pone\ applies any acceptable evaluator $f$ to \ptwo\ we have that $\ABR(G)=\ABR\left(G_f\right)$.

Unfortunately, Theorem~\ref{thm:pgbad} demonstrates that whenever $\ep>0$, even a single communication failure can cause significant harm to  emergent behavior in a game.
Note that Theorem~\ref{thm:pgbad} uses the optimistic form of the quality measure from~\eqref{eq:Qopt} with $\mathcal{E}={\rm ABR}$; that is, it evaluates the best state in any recurrent class (of the best-reply proces) of the reduced game against the worst state in any recurrent class of the nominal game.

\begin{theorem} \label{thm:pgbad}
For any $\ep>0$, let $\gee_\ep^{\rm PG}$ be the set of potential games in which Player~2 is $\ep$-inconsequential to Player~1.
There exists a game $G\in\gee_\ep^{\rm PG}$ such that if Player~1 applies any acceptable evaluator $f$ to Player~2, it holds that
\begin{equation}
Q^-_{\ABR}(G,f) \leq  \ep.  \label{eq:pgbad}
\end{equation}
\end{theorem}

That is, losing information about another agent (even an inconsequential one) can have devastating consequences.
Note that Theorem~\ref{thm:pgbad} even allows \pone\ to select the evaluator $f$ \emph{after} the pathological game is realized, indicating that even knowledge about the particular game instance is not sufficient to allow an agent to select an effective evaluator.
Theorem~\ref{thm:pgbad} leads to the following immediate corollary regarding the performance of the class of potential games as a whole:

\begin{corollary} \label{cor:pgbad}
For any $\ep>0$, let $\gee_\ep^{\rm PG}$ be the set of potential games in which Player~2 is $\ep$-inconsequential to Player~1.
For any acceptable evaluator $f$ that Player~1 applies to Player~2, it holds that
\begin{equation}
Q_\ABR^-\left(\gee_\ep^{\rm PG},f\right)  \leq \ep.
\end{equation}
\end{corollary}

\subsection{Can stochastic dynamics help prevent pathologies?}

Theorem~\ref{thm:pgbad} showed that for any evaluator, a single communication failure can render the behavior of the asynchronous best-reply process arbitrarily bad; even applied to a nominally well-behaved potential game.
We ask here if this can be rectified by letting the agents update their actions using log-linear learning, but we show that it cannot.

\begin{theorem} \label{thm:sspg}
For any $\ep>0$, let $\gee_\ep^{\rm PG}$ be the set of potential games in which Player~2 is $\ep$-inconsequential to Player~1.
There exists a game $G\in\gee_\ep^{\rm PG}$ such that if Player~1 applies any acceptable evaluator $f$ to Player~2, it holds that
\begin{equation}
Q^-_{\SSL}(G,f) \leq  \ep.  \label{eq:sspg}
\end{equation}
\end{theorem}

Theorem~\ref{thm:sspg} mirrors Theorem~\ref{thm:pgbad} exactly, except here we evaluate the quality of stochastically-stable states for log-linear learning.%
\footnote{
Here, note that Theorem~\ref{thm:sspg} can easily be shown to hold for a much more general class of learning dynamics than simply log-linear learning.
In particular, it holds for any regular perturbation of our asynchronous best-reply process (see~\cite[Theorem 4]{Young1993}).
}
Again, it implies a parallel corollary for the entire class of games:

\begin{corollary} \label{cor:sspg}
For any $\ep>0$, let $\gee_\ep^{\rm PG}$ be the set of potential games in which Player~2 is $\ep$-inconsequential to Player~1.
For any acceptable evaluator $f$ that Player~1 applies to Player~2, it holds that
\begin{equation}
Q_\SSL^-\left(\gee_\ep^{\rm PG},f\right)  \leq \ep. 
\end{equation}
\end{corollary}

\section{Identical interest games are also susceptible} \label{sec:ii}

Why are potential games subject to such severe pathologies as in Theorem~\ref{thm:pgbad}?
This is partially because in a potential game, each agent's utility function is only locally aligned with the global welfare function; it may not give the agent any information about the absolute quality of a particular action, and gives the agent no information about the utility functions of other agents.
As such, it is relatively easy to construct games in which communication failures cause \pone\ to make potential-decreasing moves while ``believing'' she is ascending the potential function.

However, in identical interest games, this does not appear to be a concern: each player has access to the full welfare function, and thus knows both the relative quality of each action \emph{and} the utility functions of all other players.
Intuitively, it seems that this additional structure may be enough to prevent pathologies.
Proposition~\ref{prop:iinotasbad} shows that when considering the special case of pure Nash equilibria, identical interest games are indeed immune to the worst of the pathologies of Theorem~\ref{thm:pgbad}, provided that the $\rm max$ evaluator is applied (see Proposition~\ref{prop:eval}).
However, we also show that this positive result does not take us far:

\begin{prop} \label{prop:iinotasbad}
For any $\ep>0$, let $\gee^{\rm II}$ be the set of all identical-interest games.
For each $G\in\gee^{\rm II}$ let $a^*\in\argmax_{a\in\aaa}W(a)$.
Let $f_{\rm max}$ denote the $\max$ evaluator, and let Player~$1$ apply $f_{\rm max}$ to Player~$2$.
Then it is always true that 
\begin{equation}\label{eq:iinotbad}
a^*\in\PNE\left(G_{f_{\rm max}}\right) \hspace{.5in} \mbox{and} \hspace{.5in}\PNE\left(G_{f_{\rm max}}\right)\subseteq\PNE(G).
\end{equation}
\end{prop}

The intuition behind~\eqref{eq:iinotbad} is simple: the $\max$ evaluator can be viewed as an attempt to be optimistic; \pone\ is assuming that \ptwo\ is maximizing $U_1$.
In an identical interest game $U_1=U_2$, so this is equivalent to assuming that \ptwo\ is maximizing her own utility function $U_2$.
Thus, at a pure Nash equilibrium, other players are best-responding to each others' actions -- and the optimism of $\max$ becomes a self-fulfilling prophecy.
However, the positive nature of Proposition~\ref{prop:iinotasbad} is tenuous: a game may have many Nash equilibria -- and~\eqref{eq:iinotbad} gives no guarantee that these are optimal.
Furthermore, the second part of~\eqref{eq:iinotbad} depends strongly on the assumption that only one communication failure occurs; if \ptwo\ also applies $f_{\rm max}$ to \pone, the resulting reduced game may have many more equilibria that are not present in the nominal game.

\subsection{Noisy dynamics in identical interest games}

Proposition~\ref{prop:iinotasbad} showed that when the $\max$ evaluator is applied, a reduced identical-interest game always has at least one good pure Nash equilibrium.
Is the optimal equilibrium always a stochastically-stable state of log-linear learning?
Unfortunately, the answer is no.

Theorem~\ref{thm:ss} uses the optimistic form of the quality measure from~\eqref{eq:Qopt}; this time, we set $\mathcal{E}=\SSL$.
That is, for log-linear learning, it evaluates the best stochastically-stable state of the reduced game against the worst stochastically-stable state of the nominal game.

\begin{theorem} \label{thm:ss}
For any $\ep>0$, let $\gee_\ep^{\rm II}$ be the set of identical interest games in which Player~2 is $\ep$-inconsequential to Player~1.
There exists a game $G\in\gee_\ep^{\rm II}$ such that if Player~1 applies any acceptable evaluator $f$ to Player~2, it holds that
\begin{equation}
Q^-_\SSL(G,f) \leq  \ep. \label{eq:ss}
\end{equation}
\end{theorem}

Requiring stochastic stability is now too much -- and Theorem~\ref{thm:ss} shows that games exist for which even inconsequential communication failures induce stochastically-stable states which are arbitrarily less efficient than those of the nominal game.
That is, the high-quality Nash equilibria guaranteed to exist by Proposition~\ref{prop:iinotasbad} need not be stochastically stable.
Once again, we state the following immediate corollary:

\begin{corollary} \label{cor:ss}
For any $\ep>0$, let $\gee_\ep^{\rm II}$ be the set of identical-interest games in which Player~2 is $\ep$-inconsequential to Player~1.
For every acceptable evaluator $f$ that Player~$1$ applies to Player~$2$, it holds that
\begin{equation}
Q_\SSL^-\left(\gee_\ep^{\rm II},f\right)  \leq \ep. 
\end{equation}
\end{corollary}

\section{Limiting the harm of communication failures} \label{sec:positive}

How can a system designer mitigate the pathologies of the previous sections?
Nominally the negative results appear quite formidable, as they appear to rule out several well-behaved classes of games, as well as learning dynamics that are generally thought to provide good efficiency guarantees.
In this section we investigate more closely what is causing the pathologies of Sections~\ref{sec:pg} and~\ref{sec:ii}, and show preliminary results on how to avoid them.
To do so, we will apply the equilibrium selection properties of log-linear learning.

\subsection{Resilience to failures via an informational paradox}

Here, we investigate the cause of the identical interest pathologies of Theorem~\ref{thm:ss}.
In the example game enabling~\eqref{eq:ss}, Player 1 essentially needs a third player's ``help'' to drive the system to a low-welfare state (see the proof of Theorem~\ref{thm:ss} and Figure~\ref{fig:iibad} in the Appendix).
However, in an identical interest game, if Player~$2$ is inconsequential to Player~$1$, then Player~$2$ is universally inconsequential; that is, also inconsequential to all other players.
This suggests that if \pone\ cannot observe \ptwo's action, then perhaps other players should not either.
Theorem~\ref{thm:all} confirms this intuition; here we are using the pessimistic form of the quality metric from~\eqref{eq:Qpess}, lending strength to our result showing it can be close to $1$.

\begin{theorem} \label{thm:all}
For all $\ep\in[0,1]$, and  let $\gee_\ep^{\rm II}$ be the set of identical-interest games in which \ptwo\ is inconsequential to \pone.
Let all players other than \ptwo\ apply the $\max$ evaluator $f_{\rm \max}$ to \ptwo.
Then it holds that
\begin{equation} \label{eq:univ}
Q^+_\SSL\left(\gee_\ep^{\rm II},f_{\rm max})\right)  = 1-\ep. 
\end{equation}
\end{theorem}
\vspace{2mm}


Note that Theorem~\ref{thm:all} presents a curious paradox, showing that performance \emph{improves} when \emph{less} communication is allowed.
That is, if \ptwo's action is hidden from \emph{every} player, the performance can be dramatically better than when it is hidden from a \emph{single} player.
Though more study is needed, one possible implication of this is as follows: suppose in a multiagent system that Agent A has a high risk of losing communication with Agent B.
Theorem~\ref{thm:all} seems to suggest that if the agents' utility functions define an identical interest game, it could be desirable to {preemptively} sever communications between Agent B and \emph{all} other agents.
Naturally, without a somewhat more detailed investigation, this should not be taken explicitly as design advice -- but nonetheless it seems to warrant more research.

\subsection{Large games are more susceptible}

What drives the negative results of Theorems~\ref{thm:pgbad} and~\ref{thm:sspg} for potential games? 
In the game used to prove the foregoing theorems, the number of action profiles was conditioned on the size of $\ep$ (see proof of Theorem~\ref{thm:pgbad} in the Appendix).
When $\ep$ was very close to $0$, the proof of Theorem~\ref{thm:pgbad} required many action profiles to generate the pathology.
Is this simply an artifact of the proof technique, or is it indicative of a deeper principle?
Here, we show that there is indeed a connection between the size of a game and the degree to which communication failures can create pathologies.
This is because for small $\ep$, the game resulting from a bounded evaluator is \emph{close} to the nominal game in a formal sense defined in~\cite{Candogan2013}, provided that the number of total action profiles is small -- and thus log-linear learning selects action profiles close to the potential-maximizing states of the nominal game.

To show positive results, here we take the pessimistic form of the quality metric~\eqref{eq:Qpess}, and let $\mathcal{E}=\SSL$.
Recall that this compares the worst stochastically-stable state of the reduced game with the best of the nominal game, setting the stage for stronger positive results.

\begin{theorem} \label{thm:candogan}
For any $\ep\geq0$, let $\gee_\ep^{{\rm P}k}$ be the set of potential games in which Player~2 is $\ep$-inconsequential to Player~1 and such that $|\aaa |\leq k$.
For every bounded acceptable evaluator $f$, it holds that
\begin{equation}
Q_\SSL^+\left(\gee_\ep^{{\rm P}k},f\right)  \geq \max\left\{0, 1 - 8\ep(k-1)\right\}. 
\end{equation}
\end{theorem}
\noindent That is, if $\ep$ is small relative to the number of action profiles in $G$,   communication failures cause limited harm.

\section{A certificate for safe proxy payoffs: coarse potential alignment} \label{sec:cert}
Though this work has shown that in general no straightforward method exists for computing good  proxy payoffs from the nominal game payoffs, it would be valuable to develop tools which could certify a set of proxy payoffs as safe for a particular setting.
In this section we propose one such certificate.
This section requires the following definition: a \emph{best-reply path} is a sequence of action profiles $\{a^1,a^2,\ldots,a^m\}$ such that for each $k\in\{1,\ldots,m-1\}$,
\begin{enumerate}[i)]
\item $a^{k+1} = (a_i,a^k_{-i})$ for some agent $i\in\mathcal{I}$ with $a_i\neq a_i^k$, and
\item $a_i^{k+1} \in B_i(a^k)$.
\end{enumerate}
That is, each successive action profile differs from the previous in the action of a single agent, and the updating agent chooses a best response.
In a potential game, it is known that all best-reply paths terminate at a pure Nash equilibrium.
A \emph{weakly-acyclic game under best replies} is a generalization of a potential game in which for every joint action profile $a\in\aaa$, there exists a best-reply path $\{a^1,a^2,\ldots,a^{m}\}$ where $a^1=a$ and $a^m$ is a Nash equilibrium.

Many simple learning rules are known that converge almost surely to Nash equilibria in weakly-acyclic games under best replies; in particular, this is true of our asynchronous best reply process defined in Section~\ref{sec:model}.
Thus, when one of these games has a \emph{unique} pure Nash equilibrium, this equilibrium may be considered highly likely to arise in game play~\cite{Young1993,Marden2012}.

Our certificate pertains to a setting in which the nominal game is a potential game with a unique pure Nash equilibrium.
In these games, pathologies could be constructed because a single player's payoffs could be specified so that any proxy payoffs would cause that player to best-respond \emph{against} the potential gradient.
To rule out such pathologies, proxy payoffs must be appropriately aligned with the potential function.
In the following theorem, we give one such characterization of ``appropriately aligned.''

Suppose that \pone\ cannot know \ptwo's action.
Given a potential game $G$ with potential function $W$, let the \emph{reduced} potential function $\tilde{W}$ be
\begin{equation}
\tilde{W}\left(a_1,a_{-12}\right) = \max\limits_{a_2\in\aaa_2} W\left(a_1,a_2,a_{-12}\right).
\end{equation}
That is, $\tilde{W}$ is the nominal potential function with the ${\rm maximum}$ evaluator applied to \ptwo.
Using this definition, we can state the following result, which holds even if we do not require inconsequentiality.

\begin{prop} \label{prop:coarse}
Let $G$ be a potential game with $n\geq3$ players and a unique pure Nash equilibrium $a^*$.
If \pone\ is assigned proxy payoff functions $\tilde{U}_1:\aaa_1\times\aaa_{-12}\rightarrow \mathbb{R}$ satisfying
\begin{equation}
\argmax\limits_{a_1\in\aaa_1}\tilde{U}_1\left(a_1,a_{-12}\right) \subseteq \argmax\limits_{a_1\in\aaa_1}\tilde{W}_1\left(a_1,a_{-12}\right),
\end{equation}
then the reduced game $\tilde{G}$ associated with $\tilde{U}$ is weakly-acyclic under best replies and has a unique pure Nash equilibrium $a^*$.
\end{prop}

A consequence of this proposition is that for every potential game with a unique pure Nash equilibrium, there do exist safe proxy payoffs.
However, these proxy payoffs may not be computable.
That is, it is not clear why a player would have access to the potential function to be able to compute these proxy payoffs, and in an arbitrary potential game, \pone's utility function need not contain enough information about the potential function for this to be a valid approach.
This highlights a crucial issue: in a potential game, despite the fact that individual players' utility functions are locally aligned with the potential function, even slight perturbations of these utility functions can essentially discard the information that is required to ascend the potential function.

\section{Conclusions}

Much work remains to be done here.
Though the negative results (e.g., of Theorem~\ref{thm:pgbad}) appear quite severe, this paper contains some encouraging results as well that leave open several important questions for future research.
For instance, all results were reported for the limited case that a single agent loses communication with a single other agent.
It seems likely that our results would only be worsened with more complex cases of information loss, but future work will shed light on this.
Another important question is the role of memory in learning algorithm design: here we studied only memoryless learning algorithms in which agents choose actions only as a function of current payoffs.
Particularly in the case of identical interest games, allowing agents to choose actions as a function of historical payoffs could conceivably provide some benefits.

\bibliographystyle{ieeetr}
\bibliography{../library/library}

\section*{APPENDIX: Proofs}

\subsubsection*{Proof of Proposition~\ref{prop:eval}}
To see that each satisfies Definition~\ref{def:eval}, let $S\in\mathbb{R}^k$ and $S'\in\mathbb{R}^k$ satisfy the first assumption of Definition~\ref{def:eval}.
Arrange $S$ and $S'$ in ascending order and denote the $i$-th element of $S$ and $S'$ as $s_i$ and $s_i'$, respectively so that $\min_{s\in S}=s_1$ and $\max_{s_\in S}=s_k$.
Thus,
\begin{align}
f_{\rm sum}(S) 	= \sum_{i=1}^k s_i 
			\ >   \sum_{i=1}^k s'_i   
		 	\ > \ f_{\rm sum}(S').
\end{align}
Since $f_{\rm sum}$ satisfies Definition~\ref{def:eval}, it must be true that $f_{\rm mean}$ does as well.
To see that $f_{\rm max}$ and $f_{\rm min}$ satisfy Definition~\ref{def:eval}, simply note that $f_{\rm max}(S) = s_k > s'_k = \ f_{\rm max}(S')$ and $f_{\rm min}(S) = s_1 > s'_1 = \ f_{\rm min}(S')$.
For these evaluators, the second axiom of Definition~\ref{def:eval} is obvious.%
\hfill\QED

\vspace{4mm}
\subsubsection*{Proof of Theorem~\ref{thm:pgbad}}
We will construct a potential game $G\in\gee^{\rm PG}_\ep$ with a unique Nash equilibrium (and thus a unique efficient recurrent class of the asynchronous best reply process) and show that for any $f$ its reduced variant $G_{f}$ is a weakly-acyclic game under best replies (see Section~\ref{sec:cert} for definition) with a unique Nash equilibrium with welfare within $\ep$ of 0.
Standard results in learning theory then imply that the Nash equilibrium of the reduced game is the unique recurrent class of the asynchronous best reply process, completing the proof.

Our constructed game has $3$ players. 
Let $M\in\mathbb{N}$ be the positive integer satisfying $1/\ep-8\leq M<1/\ep-7$.%
\footnote{Strictly, we also require $\ep<1/7$; for larger $\ep$, similar examples can be constructed to show the same bound but we omit these for reasons of space and because the result is considerably more interesting for small $\ep$.}
Player 1 has actions $\aaa_1 = \{0,1, \ldots,M+1\}$; Player 2 has actions $\aaa_2 = \{0,1,2\}$; Player 3 has actions $\aaa_3 = \{0,1, \ldots,M\}$.
The game is built of $M+1$ two-player games in which the action of Player $3$ selects which game is played between Players~$1$ and~$2$.
We refer to the actions of Players~$1$,~$2$, and~$3$ as ``rows,'' ``columns,'' and ``levels,'' respectively.

The payoff matrix which comprises Level 0 (that is, $a_3=0$) is depicted in Figure~\ref{fig:pg3}.
The two matrices in Figure~\ref{fig:pg3} depict payoffs and potential values resulting from the actions of Players~$1$ and~$2$ when $a_3=0$.
For each $(a_1,a_2)$, the upper matrix represents the value of Player $1$'s payoffs $U_1(\cdot)$; the lower matrix depicts the value of the potential function $W(\cdot)$.
Let Players 2 and 3 have payoffs equal to $W$ so for any $a$, $U_2(a)=U_3(a)=W(a)$.
For the action profiles not depicted for $a_1>2$, the payoffs and potential are equal to those when $a_1=2$.
That is, for any $m>2$, $U_1(m,a_2,0)=U_1(2,a_2,0)$ and $W(m,a_2,0)=W(2,a_2,0)$.

For $a_3>0$, consider the matrices in Figure~\ref{fig:pg-generic}.
Note that these matrices are similar to those for $a_3=0$; $u_1(a_1,a_2,k)=u_1(0,a_2,0)$ for $a_1\in\{k,k+1\}$ and $W(a_1,a_2,k)=W(0,a_2,0)-k\ep$.
They each contain the additional row $k-1$, which is simply the payoffs/potential from row $k$ plus $\ep/2$. 
For both , all rows not depicted are identical to row $k+2$.

In the nominal game, $a^{\rm ne}=(0,0,0)$ is a unique pure Nash equilibrium with potential $W(a^{\rm ne})=1$.
This can be proved by induction; the base case is depicted in Figure~\ref{fig:pg3}, which contains only $(0,0,0)$ as a Nash equilibrium.
If there is another pure Nash equilibrium, it must be associated with some $a_3>0$.
For the inductive step, consider $a_3=k$ as in Figure~\ref{fig:pg-generic}.
Here, the only possible Nash equilibrium is $(k-1,0,k)$.
At  this action profile, Player 3 has a payoff of $1+\ep/2-k\ep$, but he can deviate to $a_3=k-1$ (i.e., the next-lower level) to obtain an improved payoff of $1-k\ep+\ep$, so this cannot be a Nash equilibrium.
Therefore, $a^{\rm ne}=(0,0,0)$ is unique, so all better-reply paths in $G$ terminate at $a^{\rm ne}$.

\begin{figure}
\vspace{1mm}
\begin{center}
 \setlength{\extrarowheight}{2pt}
    $\begin{tabu}{cc|c|c|c|}
      & \multicolumn{1}{c}{U_1(\cdot)} & \multicolumn{3}{c}{\text{Player $2$}}\\
      & \multicolumn{1}{c}{} & \multicolumn{1}{c}{0}  & \multicolumn{1}{c}{1} & \multicolumn{1}{c}{2} \\\cline{3-5}
      \multirow{3}*{\text{Player $1$}}  
      & 0			& 1			& 1-3\ep		& 1-6\ep 	\\\cline{3-5}
      & 1 			& \shd1-2\ep 	& 1+\ep		& 1-5\ep 	\\\cline{3-5}
      & 2 			& 0			& 4\ep		& -2\ep		\\\cline{3-5}
\end{tabu}$
\hspace{.5in}
    $\begin{tabu}{cc|c|c|c|}
      & \multicolumn{1}{c}{W(\cdot)} & \multicolumn{3}{c}{\text{Player $2$}}\\
      & \multicolumn{1}{c}{} & \multicolumn{1}{c}{0}  & \multicolumn{1}{c}{1} & \multicolumn{1}{c}{2} \\\cline{3-5}
      \multirow{3}*{\text{Player $1$}}  
      & 0			& 1 			& 1-7\ep 		& 1-4\ep 	\\\cline{3-5}
      & 1 			&\shd 1-2\ep 	& 1-3\ep		&1-3\ep 	\\\cline{3-5}
      & 2 			& 0 			& 0 			& 0		\\\cline{3-5}
\end{tabu}$
\caption{\label{fig:pg3}Left: Player 1 payoff function $u_1(\cdot)$ for game used to prove Theorem~\ref{thm:pgbad}. 
Right: Potential function $W$ values (and Player 2/3 payoffs) for the same game. 
Both are depicted with Player 3 playing action $0$.
When Player 1 applies an acceptable evaluator to \ptwo, he prefers Action $1$ to Action $0$.
When Player $1$ plays action $1$, Player $2$'s best response is to play action $0$, making $(1,0,0)$ the only pure Nash equilibrium in this simplified game.
See Figure~\ref{fig:pg-generic} for a depiction of the game's payoffs for action profiles when Player $3$ is playing $a_3>0$.}
\end{center}
\vs\vs\vs\vs\vs
\end{figure}

Now, note that Player 2 is $6\ep$-inconsequential to Player 1.
Let Player 1 apply an acceptable evaluator $f$ to Player 2 to obtain the reduced game $G_{f}$.
Considering Figure~\ref{fig:pg-generic}, let $S_{k-1}$, $S_{k}$, $S_{k+1}$, and $S_{k+2}$ denote the depicted rows in Player 1's payoff matrix.
Order each row nondecreasing and match elements so Definition~\ref{def:eval} implies that $f(S_{k+1})>f(S_{k-1})>f(S_{k})>f(S_{k+2})$ whenever $M<1/\ep-3$.
Thus, for any $a_2\in\aaa_2$, Player 1's reduced payoff function $\tilde{U}_1(a)$ satisfies
$$
\tilde{U}_1(k+1,a_2,k)>\tilde{U}_1(k-1,a_2,k)>\tilde{U}_1(k,a_2,k).
$$
That is, when Player 3 is playing $k$, Player $1$'s best response is $a_1=k+1$, regardless of the action of Player 2.
Then, Player 2's best response to $a_1=k+1$ is to choose $a_2=0$, to obtain the gray-shaded action profile in Figure~\ref{fig:pg-generic}.

At this action profile $(k+1,0,k)$, Player $3$ can improve his payoff from $1-2\ep-k\ep$ to $1-\ep-k\ep$ by deviating to action $a_3=k+1$ (i.e., ``moving up'' one level).
Thus, all action profiles have a best-reply path which terminates at the unique Nash equilibrium $\tilde{a}^{\rm ne}=(M+1,0,M)$, with potential $W(\tilde{a}^{\rm ne})=1-2\ep-M\ep$.
That is, $G_{f}$ is a weakly-acyclic game under best replies.
Because $M\geq 1/\ep-8$, 
\begin{equation}
W(\tilde{a}^{\rm ne}) \leq1-2\ep-\left({1}/{\ep}-8\right)\ep = 6\ep.
\end{equation}
Since Player~$2$ is $6\ep$-inconsequential to Player~$1$, and the potential maximum is $1$, the theorem is proved.%
\hfill\QEDopen
%

\begin{figure}
\vspace{1mm}
\begin{center}
 \setlength{\extrarowheight}{2pt}
    $\begin{tabu}{cc|c|c|c|}
      & \multicolumn{1}{c}{U_1(\cdot)} & \multicolumn{3}{c}{\text{Player $2$}}\\
      & \multicolumn{1}{c}{} & \multicolumn{1}{c}{0}  & \multicolumn{1}{c}{1} & \multicolumn{1}{c}{2} \\\cline{3-5}
      \multirow{4}*{\text{Player $1$}}  
      & k-1 			& 1+\frac{\ep}{2}	& 1-2.5\ep 	& 1-5.5\ep		\\\cline{3-5}
      & k			& 1				& 1-3\ep		& 1-6\ep 		\\\cline{3-5}
      & k+1 			& \shd1-2\ep 		& 1+\ep		& 1-5\ep 		\\\cline{3-5}
      & k+2 			& k\ep			& 4\ep+k\ep 	& -2\ep +k\ep	\\\cline{3-5}
\end{tabu}$
\hspace{0.5in}
    $\begin{tabu}{cc|c|c|c|}
      & \multicolumn{1}{c}{W(\cdot)} & \multicolumn{3}{c}{\text{Player $2$}}\\
      & \multicolumn{1}{c}{} & \multicolumn{1}{c}{0}  & \multicolumn{1}{c}{1} & \multicolumn{1}{c}{2} \\\cline{3-5}
      \multirow{3}*{}  
      & k-1 			& 1+\frac{\ep}{2}-k\ep 		& 1-6.5\ep-k\ep 		& 1-3.5\ep-k\ep	\\\cline{3-5}
      & k			& 1 	-k\ep				& 1-7\ep 	-k\ep		& 1-4\ep-k\ep 		\\\cline{3-5}
      & k+1 			&\shd 1-2\ep 	-k\ep		& 1-3\ep	-k\ep		&1-3\ep - k\ep 		\\\cline{3-5}
      & k+2 			& 0 						& 0 					& 0				\\\cline{3-5}
\end{tabu}$
\caption{Left: Player 1 payoff function $U_1(\cdot)$ for game used to prove Theorem~\ref{thm:pgbad}. 
Right: Potential function $W$ values (and Player 2/3 payoffs) for the same game. 
Both are depicted with Player 3 playing action $k>0$.
All rows not depicted have the same payoffs (or potential values) as row $(k+2)$.
Here, given this action of Player 3, the nominal game has a unique Nash equilibrium at $(k-1,0)$.
When Player 1 applies an acceptable evaluator to \ptwo, he prefers Action $k+1$ to Action $k$; Player $2$'s best response to this is to play action $0$, making the shaded action profile $(k+1,0,k)$ the only pure Nash equilibrium in this simplified game.
Because the full game is staggered, at the action profile $(k+1,0,k)$, Player 3 has an incentive to switch to action $k+1$.}
\label{fig:pg-generic}
\end{center}
\vs\vs\vs
\end{figure}

\vspace{4mm}
\subsubsection*{Proof of Theorem~\ref{thm:sspg}}
Let $G$ be the game specified in the proof of Theorem~\ref{thm:pgbad}; for every acceptable evaluator $f$, $G_{f}$ is a weakly-acyclic game under best replies with a unique equilibrium $a^\dagger$ satisfying $W(a^\dagger)<\ep$.
Thus, $\ABR(G_f)=\{a^\dagger\}$.
For log-linear learning, we may apply~\eqref{eq:ssinabr} to conclude that $\SSL(G)=\{a^\dagger\}$, obtaining the proof.
\hfill\QEDopen

\vspace{4mm}
\subsubsection*{Proof of Proposition~\ref{prop:iinotasbad}}
Let $G$ be any identical interest game and let $f=\max$ so that $G_{f}$ is the reduced variant.
Let $a^*\in\argmax_{a\in\aaa} W(a)$; this is both the maximum-potential action profile and since the game is identical-interest, $a^*\in\PNE(G)$.
First we show that $a^*\in\PNE(G_f)$.
Since $W(a^*)\geq W(a)$ for any $a\in\aaa$, the fact that $f=\max$ implies that for \pone\ and any $a_1\in\aaa_1$,
\begin{equation}
\tilde{U}_1\left(a_1^*,a^*_{-1}\right) = W(a^*) \geq \tilde{U}_1\left(a_1,a^*_{-1}\right),
\end{equation}
where the inequality stems from the fact that $W(a^*)$ (and thus $U_1(a^*)$) is maximal with respect to any other action profile.
Thus, player $1$ cannot unilaterally deviate and improve utility.
Since $a^*$ is an equilibrium of $G$, every player $j$ with is already playing a best response to $a^*$.
Thus, $a^*$ is a pure Nash equilibrium of $G_{f}$, implying~\eqref{eq:iinotbad}.

Next, let $\bar{a}\in{\rm PNE}(G_{f})$; we wish to show that $\bar{a}\in{\rm PNE}(G)$.
Since all other players $j\neq 1$ know \pone's action, they can best-respond with respect to the true $W$.
Thus,
\begin{align}
\tilde{U}_1(\bar{a}_1,\bar{a}_{-1}) 	
							&= U_1(\bar{a}_1,\bar{a}_{-1}). \label{eq:ineq1}
\end{align}
Since $f=\max$, for all $a_1'\neq\bar{a}_1$ we have
\begin{align}
\tilde{U}_1({a}_1',\bar{a}_{-1}) 	
							&\geq U_1(a_1',\bar{a}_{-1}). \label{eq:ineq2}
\end{align}
Combining~\eqref{eq:ineq1} and~\eqref{eq:ineq2} with the fact that $\bar{a}$ is a Nash equilibrium of $G_{f}$, we have for all $a_1'\neq\bar{a}_1$ that
\vs
\begin{align}
U_1(\bar{a}_1,\bar{a}_{-1}) 	
						&\geq U_1(a_1',\bar{a}_{-1}).
						\vs
\end{align}
That is, \pone\ has no incentive to deviate from $\bar{a}$ in the nominal game.
Since every other has the same payoffs in the nominal game as in the reduced game, $\bar{a}$ must be an equilibrium of the nominal game as well.
\hfill\QEDopen

%

\begin{figure*} 
\vspace{1mm}
\begin{center}
\includegraphics[scale=0.35]{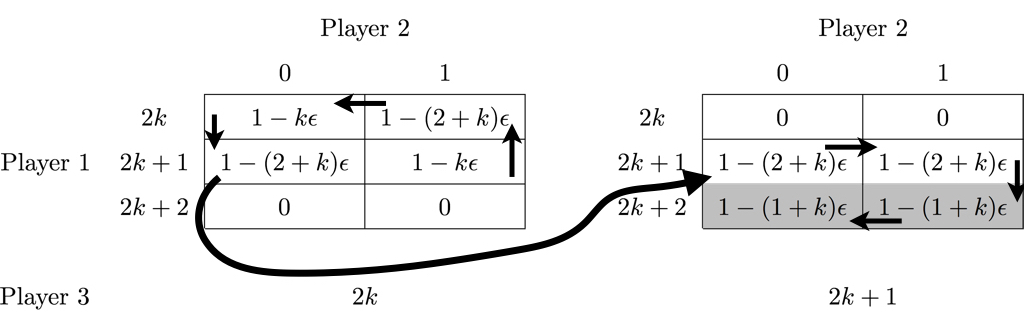}
\vs\vs\vs\vs
\caption{
Generic modular block for the identical-interest game described in the proof of Theorem~\ref{thm:ss}, displaying best-reply path resulting when Player~$1$ applies any acceptable evaluator to Player~$2$.
If \pone\ applies any acceptable evaluator to \ptwo, note that he will be indifferent between actions $2k$ and $2k+1$ when Player $3$ is playing $2k$ (i.e., the left matrix).
Thus, there is a best-reply path from every action profile depicted here to $\aaa^\dagger_k$.
Once in $\aaa^\dagger_k$, a best-response process cannot escape except by Player $3$ incrementing his action to $2k+2$. }
\label{fig:iibad}
\end{center}
\vspace{-5mm}
\end{figure*}

\vspace{4mm}
\subsubsection*{Proof of Theorem~\ref{thm:ss}}
We shall construct a $3$-player game that exhibits the pathology described in Theorem~\ref{thm:ss}.
For $\ep>0$, let $M\in\mathbb{N}$ be the positive integer satisfying $1/\ep-3\leq M<1/\ep-2$.%
\footnote{
We also require $\ep<1/3$.
Note that this is without loss of generality: when $\ep'>\ep$, it is true that $\ep'$-inconsequentiality implies $\ep$-inconsequentiality.
}
In this game, $\aaa_1=\left\{0,1,\ldots,2M+2\right\}$, $\aaa_2=\{0,1\}$, and $\aaa_3 = \left\{0,1,\ldots,2M+1\right\}$.

The game is comprised of $M$ fundamental blocks indexed by $k\leq M$, a generic one of which is depicted in Figure~\ref{fig:iibad}. 
Given a block $k$, Player $3$ has two actions: $2k$ and $2k+1$, depicted in Figure~\ref{fig:iibad} as the left and right payoff matrices, respectively.
Note that in each block, the potential-maximizing states are both in the left matrix: $(2k,0,2k)$ and $(2k+1,1,2k)$, each with a potential of $1-k\ep$.
Since these states' potential are maximal when $k=0$, the potential-maximizing states of the overall game are $(0,0,0)$ and $(1,1,0)$ with potential of~$1$.
Thus, these are also the only stochastically-stable states of log-linear learning in the nominal game.

Player $2$ is $2\ep$-inconsequential to Player $1$; let Player $1$ apply any acceptable evaluator $f$ to Player $2$.
Now, when $a_3=2k$ (left matrix), Definition~\eqref{def:eval} implies that $f(1-k\ep,1-(2+k)\ep)=f(1-(2+k)\ep,1-k\ep)$, so Player $1$ receives equal payoff for actions $2k$ and $2k+1$.
Nevertheless, when $a_3=2k+1$ (right matrix), Player $1$ still strictly prefers action $2k+2$ to action $2k+1$.
To prove Theorem~\ref{thm:ss}, we will show that for any $k\leq M$, there is a best-reply path from every action profile to the gray-shaded action profiles in Figure~\ref{fig:iibad}, but that the only best-reply path \emph{leaving} those action profiles has Player~$3$ incrementing his action to $2k+2$ (thus pushing the system state into the next-higher block, where the process repeats).

Let $k<M$, and let the gray-shaded action profiles in Figure~\ref{fig:iibad} be denoted $\aaa^\dagger_k=\left\{(2k+2,0,2k+1),\right.$ $\left.(2k+2,1,2k+1)\right\}$.
Let $a$ be an action profile such that $a_3\in\{2k,2k+1\}$ (i.e., $a$ is in block $k$).
If $W(a)=0$, note that $\aaa^\dagger_k$ can be reached in one step; either by Player~$3$ deviating to action $2k+1$ or by Player~$1$ deviating to action $2k+2$.
On the other hand, if $W(a)>0$, then $a$ lies on the best-reply path depicted in Figure~\ref{fig:iibad}.
That is, for every $a$, there is a best-reply path from $a$ to $\aaa^\dagger_k$.

Next, note that a deviation by Player $2$ cannot escape $\aaa^\dagger_k$, and Player $1$ strictly prefers the states in $\aaa^\dagger_k$ to any other states that he can reach with a single deviation.
Thus, the only way for a best-reply path to escape $\aaa^\dagger_k$ is for Player $3$ to increment his action to $a_3=2k+2$; when $k<M$, this can occur  when the state is $(2k+1,1,2k)\in\aaa^\dagger_k$.

Thus, for every action profile in block $k<M$, there is a best-reply path leading to an action profile in block $k+1$.
When $k=M$, there is a best-reply path leading to an action profile in $\aaa^\dagger_M$, but no best-reply path \emph{leaves} $\aaa^\dagger_M$.
This implies that $\ABR(G) = \aaa^\dagger_M$.
We then apply~\eqref{eq:ssinabr} to obtain $\SSL(G)\subseteq\aaa^\dagger_M$.
By the definition of $M$, it follows for any $a\in\SSL(G)\subseteq\aaa_M^\dagger$ that $W(a)\leq2\ep$, proving the theorem (since Player $2$ is $2\ep$-inconsequential).%
\hfill\QEDopen

\vspace{4mm}
\subsubsection*{Proof of Theorem~\ref{thm:all}}

Suppose that \ptwo\ is $\ep$-inconsequential to \pone; because this is an identical interest game, this implies that Player 2 is $\ep$-inconsequential to all players $\mathcal{I}\setminus\{2\}$.
If all players other than 2 ignore the actions of Player 2 by applying an acceptable evaluator $f$ to their utility functions, then the action choice of Player 2 has no effect on the decisions of players other than 2.
This means that the reduced game can be analyzed as an $(n-1)$-player identical interest game, and Player 2 can be modeled as a simple optimizer.

Let all players $\mathcal{I}\setminus\{2\}$ apply the maximum evaluator $f=\max$, so that the reduced payoff functions $\tilde{U}$ and corresponding reduced potential function $\tilde{W}$ for each player $i\in\mathcal{I}\setminus\{2\}$ are given by
$
\tilde{W}(a) = \tilde{U}_i(a) = \max_{a_2\in\aaa_2}U_i\left(a_{-2},a_2\right).
$
The reduced game, denoted $\tilde{G}$, 
will be an identical-interest game played between all players in $\mathcal{I}\setminus\{2\}$.

Let $\tilde{\aaa}^*=\argmax_{a\in\aaa}\tilde{W}(a)$ be the set of potential-maximizing states of $\tilde{G}$; because $\tilde{G}$ is an identical-interest game, $\tilde{\aaa}^*$ must also be the set of stochastically-stable states of log-linear learning.
All stochastically-stable states for the reduced game $G_{f}$ must also be stochastically-stable for $\tilde{G}$; thus, to prove the theorem, it will suffice to establish a tight lower bound on potential for states in $\tilde{\aaa}^*$.

Since the players are applying the maximum evaluator, it must be the case for any state $\tilde{a}\in\tilde{\aaa}^*$ that there exists some $a_{2}$ such that $W(a_2,\tilde{a}_{-2})=1$. 
Thus, $\ep$-inconsequentiality provides the lower bound of
$W(\tilde{a}) \geq W(a_2,\tilde{a}_{-2}) - \ep$.
The game in Figure~\ref{fig:iibad} with $k=0$ shows the bound to be tight.
\hfill\QEDopen

\vspace{4mm}
\subsubsection*{Proof of Theorem~\ref{thm:candogan}}
The proof of Theorem~\ref{thm:candogan} relies on the following definition:
\begin{definition}[Candogan et al.,~\cite{Candogan2013}]
Let $G$ and $\hat{G}$ be two games with players $\mathcal{I}$, action set $\aaa$, and utility functions $\{U_i\}_{i\in\mathcal{I}}$ and $\{\hat{U}_i\}_{i\in\mathcal{I}}$ respectively.
Let $\Delta_i(G,\hat{G},a,a_i')$ denote
\begin{equation}
\left|
\left(U_i(a_i,a_{-i})- U_i(a'_i,a_{-i})\right) - 
\left(\hat{U}_i(a_i,a_{-i})-\hat{U}_i(a'_i,a_{-i})\right)
\right|.
\end{equation}
The \emph{maximum pairwise difference (MPD)} between $G$ and $\hat{G}$ is defined as
\begin{equation}
d(G,\hat{G}) \triangleq \max\limits_{a\in\aaa,\ i\in\mathcal{I},\ a'_i\in\aaa_i}\Delta_i(G,\hat{G},a,a_i')
\end{equation}
\end{definition}

Let $G\in\gee_\ep^{{\rm P}k}$, and let $f$ be a bounded acceptable evaluator $f$ generating reduced utility function $\tilde{U}_1$.
Let $U^*_1(a_1) := U_1(a_1,a_{-1})$.
For any $a\in\aaa$ and $a_i'\in\aaa_i$, we have that
\begin{align}
\Delta_i(G,G_{f},a,a_i') 	&= \left|
U^*_i(a_i) - U^*_i(a'_i) - 
\tilde{U}^*_i(a_i) + \tilde{U}^*_i(a'_i)
\right| \nonumber \\
		&\leq \left|
U^*_i(a_i) - \tilde{U}^*_i(a_i)\right| \hs+\hs 
\left|\tilde{U}^*_i(a'_i) \hs -\hs U^*_i(a'_i) 
\right| \nonumber\\
		&\leq \ep + \ep. \nonumber
\end{align}
The first inequality is the triangle inequality; the second follows from the $\ep$-inconsequentiality of \ptwo\ and the boundedness of $f$, 
implying that $d(G,G_{f})\leq2\ep$.
Corollary 4.3 of~\cite{Candogan2013} states that the stochastically stable states of $\hat{G}$ under log-linear learning are within $4d(G,\hat{G})(|\aaa|-1)$ of the potential function maxima of $G$; applying this completes the proof of the theorem.
\hfill\QEDopen

\vspace{4mm}
\subsubsection*{Proof of Proposition~\ref{prop:coarse}}
Let $G$ and $\tilde{G}$ satisfy the assumptions of Proposition~\ref{prop:coarse}, and let $a\in\aaa$ be any action profile.
To show that $\tilde{G}$ is weakly acyclic under best replies, it suffices to construct a best-reply path from $a$ to $a^*$ for $\tilde{G}$.
Denote this best-reply path by $\{a^1,a^2,\ldots,a^m\}$, where $a^1=a$.

First, let \pone\ choose a best response; second, let \ptwo\ choose a best response.%
\footnote{
Assume without loss of generality that at least one of \pone\ and \ptwo\ are not already playing a best response when it is their chance to deviate.
}
Thus, by the definition of $\tilde{U}_1$, $a^3$ satisfies
\vs
\begin{equation}
\left( a^3_1,a^3_2 \right) \in \argmax\limits_{a_1,\ a_2} W\left(a_1,a_2,a_{-12}^1\right), \label{eq:phimax}
\vs
\end{equation}
and it must be true that $W(a^3)>W(a)$.

If $a^3$ is a pure Nash equilibrium, we are done.
Otherwise, there exists a player $j\in\{3,\ldots,n\}$ that can strictly improve its payoff with a best reply; such a deviation strictly increases the value of the potential function.
In this case, let player $j$ deviate, and then repeat the process with Players~1 and~2 as before.
In this way, since there are a finite number of action profiles and the potential function is strictly increasing along this best-reply path, it can easily be seen that a best-reply path can be found from $a$ to some Nash equilibrium $a^m$ of $\tilde{G}$.

Since $a^m$ satisfies~\eqref{eq:phimax}, it must also be a Nash equilibrium of $G$; since $G$ has a unique equilibrium, it must be that $a^m=a^*$ and the proof is obtained.%
\hfill\QEDopen

\end{document}